\documentclass{article}

\usepackage{arxiv}

\usepackage[utf8]{inputenc} 
\usepackage[T1]{fontenc}    
\usepackage{hyperref}       
\usepackage{url}            
\usepackage{booktabs}       
\usepackage{amsfonts}       
\usepackage{nicefrac}       
\usepackage{microtype}      
\usepackage{lipsum}
\usepackage{amsmath}
\usepackage{graphicx}
\usepackage{amssymb}
\usepackage{float}

\newcommand{\beginappendix}{%
        \setcounter{table}{0}
        \renewcommand{\thetable}{A\arabic{table}}%
        \setcounter{figure}{0}
        \renewcommand{\thefigure}{A\arabic{figure}}%
        \setcounter{section}{0}
        \renewcommand{\thesection}{A\arabic{section}}%
        \setcounter{equation}{0}
        \renewcommand{\theequation}{A\arabic{equation}}%
     }

\begin{document}

\begin{flushleft}
{\Large
\textbf\newline{\textbf{Cross-compensation of Zernike aberrations in Gaussian optics}}
}
\newline
\\
Jakub Czuchnowski\textsuperscript{1,\S},
Robert Prevedel\textsuperscript{1,*}
\\
\bigskip
\textsuperscript{1} Cell Biology and Biophysics Unit, European Molecular Biology Laboratory, Heidelberg, Germany
\\



\textsuperscript{\S} Collaboration for joint PhD degree between EMBL and Heidelberg University, Faculty of Biosciences, Germany
\\
\bigskip
* Correspondence to prevedel@embl.de

\end{flushleft}

\begin{abstract}
Zernike polynomials are one of the most widely used mathematical descriptors of optical aberrations in the fields of imaging and adaptive optics. Their mathematical orthogonality as well as isomorphisms with experimentally observable aberrations make them a very powerful tool in solving numerous problems in beam optics. However, Zernike aberrations show cross-coupling between individual modes when used in combination with Gaussian beams, an effect that has not been extensively studied. Here we propose a novel framework that is capable of explaining the fundamental cross-compensation of Zernike type aberrations, both in low-aberration and high-aberration regimes. Our approach is based on analysing the coupling between Zernike modes and different classes of Laguerre-Gauss modes which allows investigating aberrated beams not only on a single plane but also during their 3D propagation.

\end{abstract}

\section{Introduction}


Gaussian beams are among the most commonly encountered optical waves in modern imaging systems. It is thus crucial to further our understanding of the effects optical aberrations can exhibit on Gaussian beams. This is especially true in the recently emerging field of active aberrations correction, colloquially known as adaptive optics. Adaptive optics (AO) is an approach that uses active optical elements for a dynamic manipulation of light amplitude and phase with fine spatial resolution in order to correct for the presence of aberration and drastically increase image quality under experimental conditions \cite{Booth:14,Ji:17}. Practical implementations of AO usually require the choice of a particular orthonormal base into which experimentally measured aberrations can be decomposed. The Zernike polynomials are a frequent and popular choice since they facilitate efficient aberration correction in practice. \cite{booth2007adaptive}.

It has been shown previously that Zernike aberrations are orthogonal only in the case of plane waves and show cross coupling in the case of Gaussian beams. The effects of this cross couplings were analysed based on the Strehl ratio approach \cite{mahajan1994zernike,mahajan1995zernike}. Here we propose an alternative formulation based on analysing the effects Zernike type aberrations in inducing power coupling into higher order Laguerre-Gauss modes \cite{Mah:19,Bond:11,Czuchnowski:20} and use it to gain intuitive understanding of various model situations in AO such as iterative improvement of cross-compensating Zernike aberrations and cascading aberrations. We also extend our approach to shed light on Zernike aberration interactions in the high aberration regimes.

\section{Coupling between Zernike aberrations and Laguerre-Gauss modes in the low aberration regime}

Laguerre-Gauss beams as eigenvectors of the wave equation are inherently orthogonal and therefore do not cross-couple between each other. However it has been shown that Zernike type aberrations can induce cross coupling between different LG-beams \cite{Bond:11}. The coupling coefficient can be expressed as:

\begin{equation}
    k^{n,m}_{p,l,p',l'}=\int_S LG_{p,l}\exp(2ikZ^m_n)LG_{p',l'}^*dS
    \label{eq:knmplpl}
\end{equation}

in the low aberration regime we can approximate:

\begin{equation}
    \exp(2ikZ^m_n)=1+2ikZ^m_n
\end{equation}

which simplifies \textbf{Equation \ref{eq:knmplpl}} to an analytically solvable integral:

\begin{equation}
    k^{n,m}_{p,l,p',l'}=\int_0^{2\pi} \int_0^R LG_{p,l}LG_{p',l'}^*(2ikZ^m_n)rdrd\phi
\end{equation}

The equation can be separated into azimuthal and radial parts which can be solved independently, starting with the azimuthal part:

\begin{equation}
    \phi^{n,m}_{p,l,p',l'}=\int_0^{2\pi} e^{i\phi(l-l')}\frac{e^{im\phi}+e^{-im\phi}}{2}d\phi=\bigg[\frac{e^{i\phi(l-l'+m)}}{2i(l-l'-m)}+\frac{e^{i\phi(l-l'+m)}}{2i(l-l'-m)}\bigg]_0^{2\pi}
\end{equation}

which specifies the coupling condition: 
\begin{equation}
\label{eq:phi}
\phi^{n,m}_{p,l,p',l'} =
    \begin{cases}
      0 & \text{if $m \neq |l-l'|$}\\
      \pi & \text{if $m=|l-l'|$, \ even $Z^m_n$}\\
      sgn(l-l') i\pi & \text{if $m=|l-l'|$, \ odd $Z^m_n$}\\
      2\pi & \text{if $m=0$, \ $l=l'$}
    \end{cases}   
\end{equation}

\begin{figure}
\includegraphics[width=14cm]{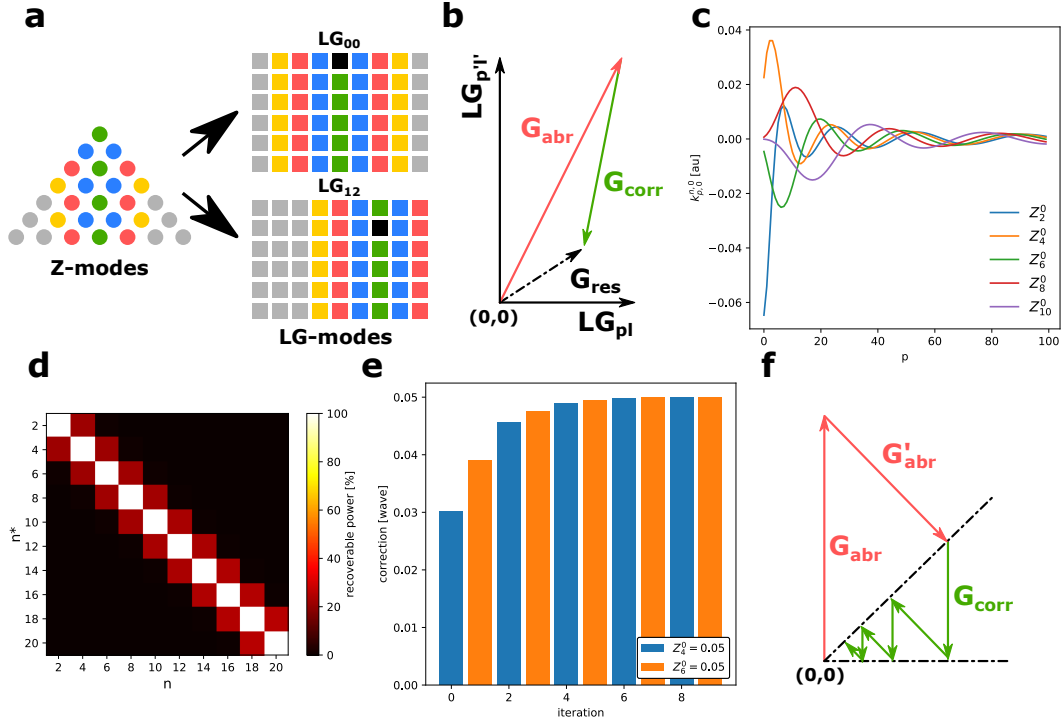}
\centering
\caption{\textbf{a.} Coupling from classes of Z-modes to classes of LG-modes depending on which LG beam is assumed as the fundamental beam. \textbf{b.} Schematic representation of aberrated beams ($G_{abr}$) as vectors in the LG space showing the possibility of cross compensation by non-orthogonal aberrated beams ($G_{corr}$). \textbf{c.} Amplitude of coupling coefficients into different LG-modes induced by Zernike aberrations of the spherical aberration class. \textbf{d.} Quantification of the cross-compensation between different modes in the spherical aberration class ($Z_n^0$), expressed as the percentage of power out-coupled from the fundamental Gaussian mode that can be recovered. \textbf{e.} Iterative adaptive optics correction for two cross coupling aberrations. \textbf{f.} Geometric interpretation of iterative adaptive optics correction for two cross coupling aberrations. The iterative improvement is trapped between the two dotted lines which are the orthogonals of the aberration vectors.} 
\label{fig:1}
\end{figure}

This means that efficient coupling between modes can occur only under the condition that $m=|l-l'|$. As in most imaging applications we are mainly concerned with aberrations acting on the fundamental Gaussian mode ($LG_{00}$) this simplifies the coupling condition to $m=|l'|$. This has fundamental consequences for our analysis as it provides a mapping from particular sets of Z-modes into particular families of LG-beams (\textbf{Figure \ref{fig:1}a}).

\subsection{Applications in adaptive optics}

As Zernike polynomials are frequently used as an orthonormal base for applications in adaptive optics \cite{booth2007adaptive} and aberration correction it is interesting to investigate the consequences of the cross-coupling in the context of our framework. In order to do so we first describe an AO task within our theory. 

As we operate in the LG-beam space we need to define how the effects of Zernike aberrations on the beam quality can be quantified. The most intuitive metric would be to look at the power fraction coupled into higher order LG-beams. In this general sense the goal of aberration correction could be defined as redirecting the power back into the fundamental Gaussian mode. The coupling condition ($m=|l'|$) allows us to easily map all beams aberrated by a particular class of Zernike aberrations into a subspace of LG-beam vector space:

\begin{equation}
    \mathbf{G^{n,m}}=\sum_p k^{n,m}_{p,\pm m}\mathbf{LG_{p,\pm m}}
\end{equation}

where $\mathbf{LG_{p,\pm m}}$ are understood as base vectors and $k^{n,m}_{p,\pm m}$ is the amplitude coupling coefficient \cite{Bond:11} (see \textbf{Figure \ref{fig:1}b, c}). The power coupled into higher order modes can then be simply calculated as the square of the vector length:


\begin{equation}
    P_{n,m}=||\mathbf{G^{n,m}}||^2
\end{equation}

Additionally, in the low aberration regime interactions between different Zernike aberrations translate to simple vector additions and their effect on the outcoupled power can be expressed as follows (\textbf{Figure \ref{fig:1}b}):

\begin{equation}
    P_{n*,m*}^{n,m}=||\alpha\mathbf{G^{n,m}}+\beta\mathbf{G^{n*,m*}}||^2
\end{equation}

If we assume that $\alpha$ is fixed and $\beta$ can be adjusted to attempt an AO correction the optimal reduction of outcoupled power can be calculated from geometrical relations between vectors:

\begin{equation}
    min\{P_{n*,m*}^{n,m}\}=\bigg[ \alpha||\mathbf{G^{n,m}}||\sin\zeta \bigg]^2
\end{equation}

where $\zeta$ is the angle between vectors $\mathbf{G^{n,m}}$ and $\mathbf{G^{n*,m*}}$. In case $\zeta=\pi/2$ the two vectors are orthogonal and unable to cross-compensate, otherwise the vectors can cross-compensate at least residually. We investigate the cross-coupling characteristics in the spherical aberration class ($Z^0_n$) and show that there is significant cross-compensation only between neighbouring aberrations (\textbf{Figure \ref{fig:1}d}). As the angle between vectors was calculated numerically for a finite set of LG-beam it has an uncertainty from discarding the higher order beams. We propose a simple estimation for the upper bound of this uncertainty based on vector properties (\textbf{Appendix A1}) and show that the angle is significantly different from $\pi/2$ for the considered cases.





\subsection{Correcting multiple cross-compensating aberrations}

Now we can use our framework to model a typical correction scheme in indirect adaptive optics where different Zernike corrections are applied and optimized in an independent (mode-by-mode) as well as iterative fashion. This is a particularly relevant situation in applied optics where one can see the effects of aberration cross-compensation:

\begin{equation}
    \mathbf{G^{n,m,n*,m*}}=(\alpha_1+\beta_1)\mathbf{G^{n,m}}+(\alpha_2+\beta_2)\mathbf{G^{n*,m*}}
\end{equation}

where $\alpha_1$ and $\alpha_2$ are fixed aberrations and $\beta_1, \beta_2$ are the correction coefficients which are updated iteratively. Aberration cross-compensations will cause the coefficients to evolve over several iterations until they converge at the proper values (\textbf{Figure \ref{fig:1}e}). This behaviour can be intuitively understood by looking at the geometrical interpretation of our theory (\textbf{Figure \ref{fig:1}f}). The situation for only a single fixed aberration is conceptually the same (\textbf{Appendix A2}). It is important to note here that odd ($m < 0$) and even ($m \geq 0$) Zernike aberrations do not cross-compensate, even though they couple into the same LG modes, because they are orthogonal in phase (as can be seen from \textbf{Equation \ref{eq:phi}}).

\begin{figure}
\includegraphics[width=15cm]{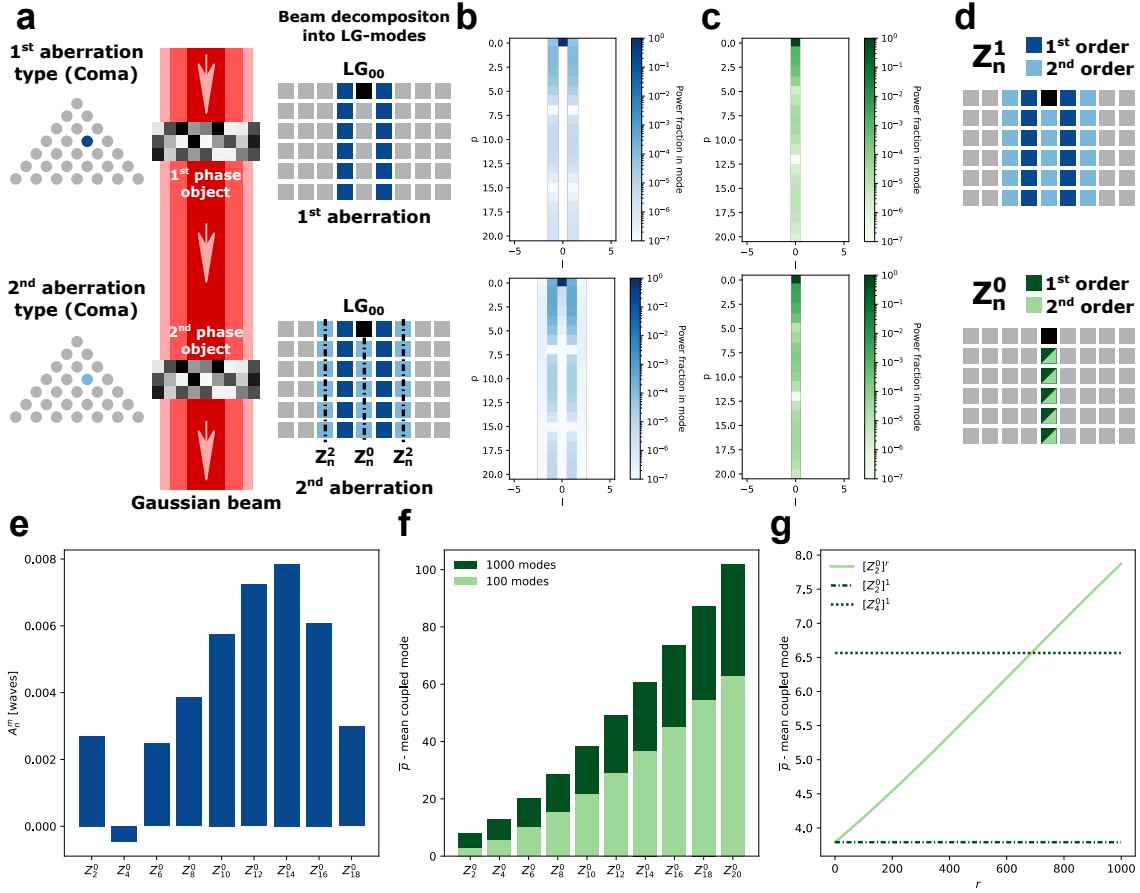}
\centering
\caption{\textbf{a.} Sequentially cascading two $Z^1_3$ aberrations (coma). \textbf{b.} Powered coupled to LG-modes for a cascading $Z^1_3$ aberration. \textbf{c.} Powered coupled to LG-modes for a cascading $Z^0_2$ aberration. \textbf{d.} Schematic showing $1^{st}$ and $2^{nd}$ order coupling into LG-modes for $Z^{\pm1}_n$-class and $Z^{0}_n$-class aberrations.  \textbf{e.} Z-mode decomposition of the residual power coupled into the $LG_{p,0}$-family caused by a cascading $Z^1_3$ aberration (\textbf{panel b}) showing possible cross-compensation between Z-classes. \textbf{f.} Mean coupled mode ($\overline{p}$) quantification for the $Z^{0}_n$-class aberrations showing that higher order $Z^{0}_n$-aberrations couple power into higher order LG-modes. \textbf{g.} Mean coupled mode ($\overline{p}$) quantification for a cascading $Z^{0}_2$ aberration showing that for higher order cascades (and consequently higher order coupling coefficients) power is being transferred into higher order LG-modes resembling the effects of higher order Z-aberrations.} 
\label{fig:2}
\end{figure}

\section{Cascading aberrations and the high aberrations regime}

The use of LG space in our framework allows us to look into aberrated beam evolution in a very straightforward manner as LG modes are eigenvectors of the wave equation. This allows us to address questions concerning effects of cascading aberrations where the evolving beam interacts twice with the same type of Zernike aberration (\textbf{Figure \ref{fig:2}a}). In this case we see that re-applying a $Z^1_3$ (coma) aberration to a $Z^1_3$-aberrated Gausian beam will cause coupling into LG-modes characteristic for the sperical aberration ($Z^0_n$) and trefoil ($Z^{\pm2}_n$) Z-mode clases (\textbf{Figure \ref{fig:2}b}). 

Conceptually cascading aberrations show similarities to the effects of Zernike aberrations in the high aberrations regime which we will now discuss. In the high aberrations regime it becomes necessary to consider higher order coupling terms which modify the coupling conditions between Z-modes and LG-modes \cite{Bond:14}:

\begin{equation}
    \exp(2ikZ^m_n)=1+2ikZ^m_n+2(ikZ^m_n)^2
\end{equation}

The resulting approximation of \textbf{Equation 1} is still solvable analytically in an analogical way \cite{Bond:14} and the resulting coupling condition from the azimutal integral is the following:

\begin{equation}
\label{eq:phi2}
\phi^{n,m,2}_{p,l,p',l'}=
    \begin{cases}
      0 & \text{if $2m \neq |l-l'|$, \ $l \neq l'$}\\
      \frac{\pi}{2} & \text{if $2m=|l-l'|$, \ even $Z^m_n$}\\
      -\frac{\pi}{2} & \text{if $2m=|l-l'|$, \ odd $Z^m_n$}\\
      \pi & \text{if $l=l'$, $m \neq 0$}\\
      2\pi & \text{if $l=l'$, $m=0$}\\
    \end{cases}   
\end{equation}

We see now that $2^{nd}$ order coupling populates the same LG-modes as a cascading aberration, which allows us to intuitively look at the Z-mode cross-compensation in the high aberration regime (\textbf{Figure \ref{fig:2}a, d}).

We observe that there are two special classes of Zernike aberrations. The $Z^0_n$-class (spherical aberrations) is special, because of the coupling from $Z^0_n$-modes is always constrained to the same class of LG-modes ($LG_{p,0}$) regardless of the strength of the aberration (\textbf{Figure \ref{fig:2}c, d}). Because of this we can define an easy metric to look at the scaling between the different orders of Z-modes and LG-modes. We show that higher order $Z^0_n$-modes couple to higher order LG-modes (\textbf{Figure \ref{fig:2}f}, see \textbf{Appendix A3} for details), and interestingly that cascading lower order aberrations ($Z^{0}_2$) over several iterations start to shift the power coupling into higher order modes (\textbf{Figure \ref{fig:2}g}). Additionally due to \textbf{Equation \ref{eq:phi2}} all Zernike aberrations will cause coupling into the $LG_{p,0}$-class if the $2^{nd}$ order is considered, facilitating possible cross-compensation between different Z-mode classes (\textbf{Figure \ref{fig:2}b, e}). Secondly, the $Z^{\pm1}_n$-class (coma) is special because it densely populates the LG-space as the aberration magnitude becomes larger, which will also facilitate possible cross-compensation. Higher order Zernike aberration classes populate the LG-space in a progressively more sparse manner thus facilitating possible cross-compensation to a lower degree.


%




\section{Discussion}

With the recent rise of popularity and applicability of adaptive optics in imaging demands a better understanding of the ways optical aberrations interact both with each other as well as with the active optical elements used for AO correction (such as deformable mirrors and spatial light modulators). As most of these devices are being used and calibrated in the Zernike aberration base the limitations of Zernike polynomials as optical aberration models need to be characterised and highlighted. In our work we proposed a novel framework based on analysing aberrated beams in the Laguerre-Gauss mode space as a useful tool for understanding the interactions of Zernike type aberrations with Gaussian laser beams. Our framework provides an intuitive explanation for common phenomena observed in experimental AO applications, namely the need to iterate over the same Zernike modes several times in order to achieve optimal correction results. Additionally, it describes the differences in the cross-coupling behaviour of Zernike aberrations between the low and high aberration regimes. Therefore our work might provide a basis to design superior correction strategies in AO or other applications based on aberrations described by Zernike basis sets.


\section*{Funding}
This work was supported by the European Molecular Biology Laboratory (EMBL).

\section*{Disclosures}

The authors declare no conflicts of interest.

\bibliographystyle{naturemag}  
\bibliography{bib}

\newpage
\clearpage

\beginappendix
{\Large\textbf{{Appendix}}}


\section{Lower and upper bound of angle approximation}

Using a geometrical vector interpretation of our theory we can easily define the lower band ($\gamma_{LB}$) and upper band ($\gamma_{UB}$) respectively where:

\begin{equation}
    \zeta_{LB}(\mathbf{G^{n,m}},\mathbf{G^{n*,m*}})=\zeta-\Delta\zeta^{n,m}-\Delta\zeta^{n*,m*}
\end{equation}

and 

\begin{equation}
    \zeta_{UB}(\mathbf{G^{n,m}},\mathbf{G^{n*,m*}})=\zeta+\Delta\zeta^{n,m}+\Delta\zeta^{n*,m*}
\end{equation}

where 
\begin{equation}
    \Delta\zeta^{n,m}=\arctan{\sqrt{\frac{1-|k_{GG}^{n,m}|^2-||\mathbf{G^{n,m}}||^2}{||\mathbf{G^{n,m}}||^2}}}
\end{equation}

Here, $k_{GG}^{n,m}$ is the amplitude coupling coefficient between an aberrated Gaussian beam and the fundamental $LG_{00}$ mode including the 2nd order correction term derived from ref. \cite{Bond:14}:

\begin{multline}
    k_{GG}^{n,m}=1+2Aik\sum_{h=0}^{\frac{1}{2}(n-m)}\frac{(-1)^h(n-h)!}{(\frac{1}{2}(n+m)-h)!(\frac{1}{2}(n-m)-h)!h!X^{\frac{1}{2}(n-2h)}}\gamma(1-h+\frac{1}{2}n,X) \\
    -2k^2A^2 \sum_{h=0}^{\frac{1}{2}(n-m)} \sum_{g=0}^{\frac{1}{2}(n-m)} \frac{(-1)^{h+g}(n-h)!X^{h+g-n}}{(\frac{1}{2}(n+m)-h)!(\frac{1}{2}(n-m)-h)!h!}\times \\
    \times \frac{(n-g)!}{(\frac{1}{2}(n+m)-g)!(\frac{1}{2}(n-m)-g)!g!}\gamma(n-h-g+1,X)
\end{multline}

where $X=2R^2/\omega^2$ describes the scaling between the Zernike radius ($R$) size and the beam radius ($\omega$), $A$ denotes the Zernike aberration amplitude, $k=2\pi/\lambda$ is the wavenumber and $\gamma(a,x)=\int^x_0 t^{a-1}e^{-t}dt$ is the lower incomplete gamma function.

\begin{figure}[h!]
\includegraphics[width=8cm]{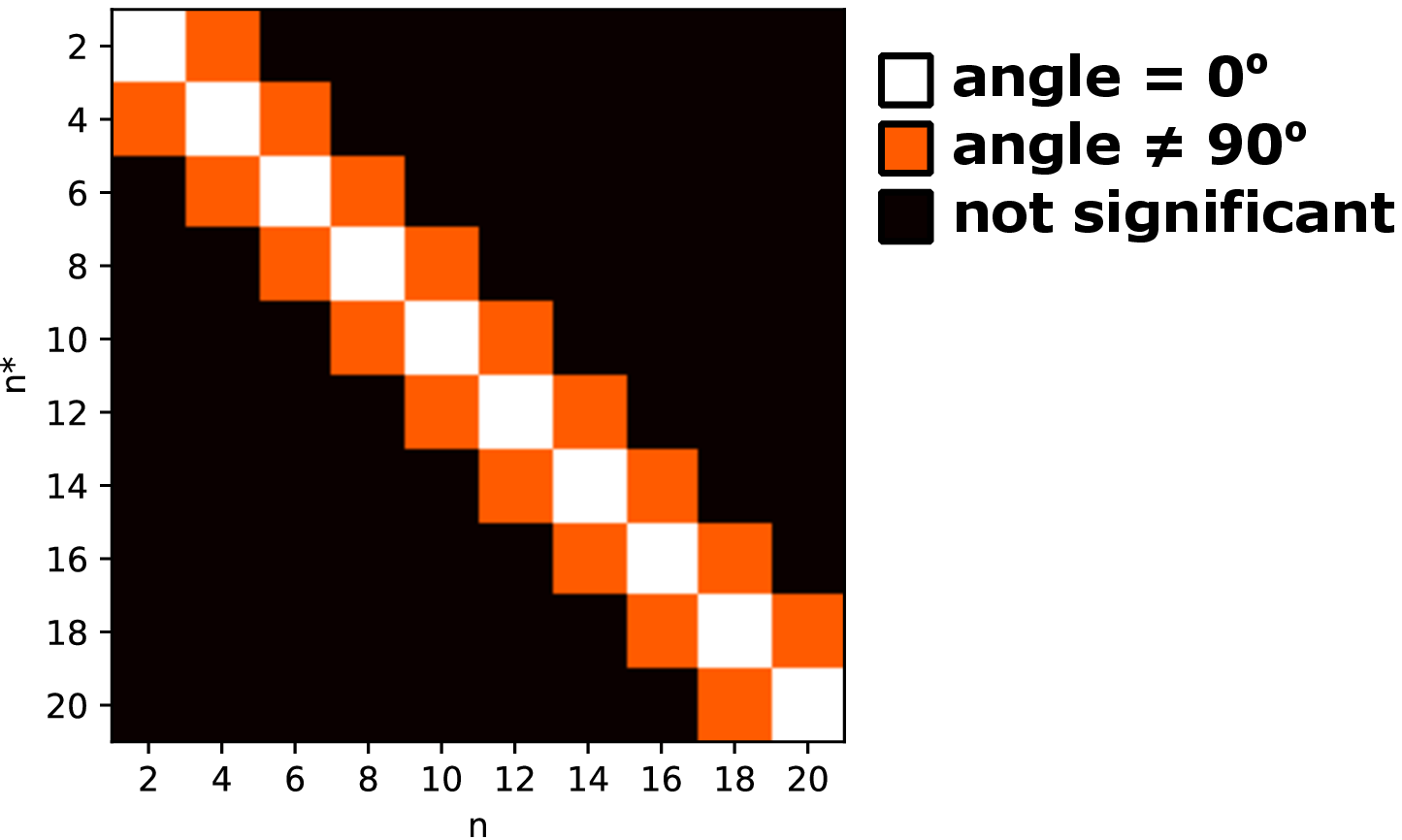}
\centering
\caption{Angles between the members of the $Z^0_n$-class (spherical) of aberrations.}
\label{fig:StatisticalSignificance}
\end{figure}

\textbf{Figure \ref{fig:StatisticalSignificance}} shows the aberrations for which the angles are significantly different form $90^o$ which nicely overlaps with the aberrations that are predicted to have a significant amount of cross-compensation (\textbf{Figure \ref{fig:1}d}). The angle is determined to be $\neq 90^o$ if $90^o \notin [\zeta_{LB},\zeta_{UB}]$, otherwise the pair of aberrations is labeled as not significant.

\section{Cross-compensation for a single fixed aberration}

\begin{figure}[h!]
\includegraphics[width=4cm]{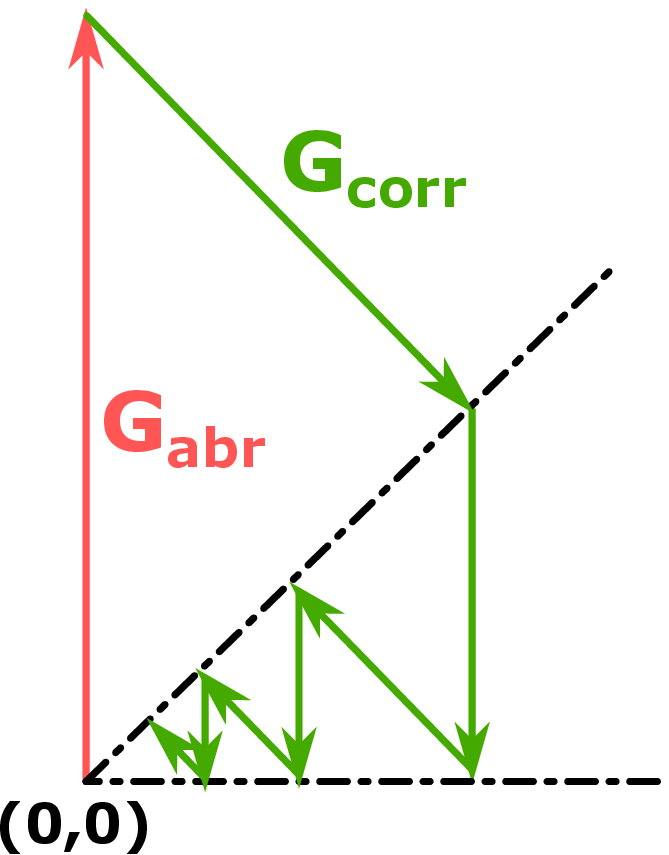}
\centering
\caption{Correction of a single aberration with two cross-compensating Zernike modes.}
\label{fig:Correction of a single aberration with two cross-compensating Zernike modes.}
\end{figure}

The situation of correcting two fixed aberrations with a set of two cross-compensating Zernike modes (\textbf{Equaiton 10} in the main text) is analogous to the situation of correcting only a single aberration:

\begin{equation}
    \mathbf{G^{n,m,n*,m*}}=(\alpha_1+\beta_1)\mathbf{G^{n,m}}+\beta_2\mathbf{G^{n*,m*}}
\end{equation}

where $\alpha_1$ is the fixed aberration and $\beta_1, \beta_2$ are the correction coefficients which are updated iteratively. \textbf{Figure \ref{fig:Correction of a single aberration with two cross-compensating Zernike modes.}} shows the geometrical interpretation of this process that is highly similar to the situation of two fixed aberrations, with one caveat being that if $\beta_1$ is updated before $\beta_2$ with only a single aberration one can achieve the perfect correction in one iteration.

\section{Mean coupled mode}

The mean coupled mode ($\overline{p}$) for the spherical aberration family ($Z^0_n$) can be defined as follows:

\begin{equation}
    \overline{p}^{n}=\sum_{p'=0}^s p'|k^{n,0}_{0,0,p',0}|^2
\end{equation}

in an ideal situation $s=\infty$ however in practice $\overline{p}^{n}$ can only be calculated for a finite number of LG modes.

\end{document}